\documentclass{article}

\usepackage{arxiv}
\usepackage[T1]{fontenc}
\usepackage{graphicx}
\usepackage{textcomp}
\usepackage{xcolor}
\usepackage{listings}
\usepackage{balance}
\usepackage{tcolorbox}
\usepackage{adjustbox}
\usepackage{siunitx}
\usepackage{subcaption}
\usepackage{bbding}
\usepackage{pifont}
\usepackage{multirow}
\usepackage{listings}
\usepackage{minted}
\usepackage{ifthen}
\usepackage{enumitem}
\usepackage{url}
\usepackage{xurl}

\setminted[java]{
xleftmargin=15pt,
framesep=2mm,
linenos=true,
breaklines=true,
tabsize=2,
encoding=utf8,
frame=none,
fontsize=\scriptsize,
escapeinside=|| 
}

\lstdefinestyle{prompt}{
    commentstyle=\color{codegreen},
    keywordstyle=\color{magenta},
    stringstyle=\color{codepurple},
    basicstyle=\ttfamily\footnotesize,
    breakatwhitespace=false,         
    breaklines=true,                 
    captionpos=b,
    keepspaces=false,               
    showspaces=false,                
    showstringspaces=false,
    showtabs=false,                  
    tabsize=2,
    literate={\Ü}{{\"U}}1
             {\ä}{{\"a}}1
             {\ö}{{\"o}}1
             {\ü}{{\"u}}1
             {\≤}{{$\leq$}}1
}

\lstdefinestyle{code}{
    backgroundcolor=\color{backcolour},   
    commentstyle=\color{codegreen},
    keywordstyle=\color{magenta},
    numberstyle=\tiny\color{codegray},
    stringstyle=\color{codepurple},
    basicstyle=\ttfamily\footnotesize,
    breakatwhitespace=false,         
    breaklines=true,                 
    captionpos=b,                    
    keepspaces=true,                 
    numbers=left,                    
    numbersep=5pt,                  
    showspaces=false,                
    showstringspaces=false,
    showtabs=false,                  
    tabsize=2
}

\newcommand{\wcircle}[1]{\ding{\numexpr171 + #1}}

\lstdefinestyle{mystyle}{
    backgroundcolor=\color{lightgray!40},   
    commentstyle=\color{green},
    keywordstyle=\color{blue},
    numberstyle=\tiny\color{gray},
    stringstyle=\color{red},
    basicstyle=\ttfamily\footnotesize,
    breakatwhitespace=false,         
    breaklines=true,                 
    captionpos=b,                    
    keepspaces=true,                 
    numbers=left,                    
    numbersep=5pt,                  
    showspaces=false,                
    showstringspaces=false,
    showtabs=false,                  
    tabsize=2
}

\begin{document}
\pagenumbering{gobble}


\title{Security Evaluation of Android apps in budget African Mobile Devices}

\author{
Alioune DIALLO \\
\textit{SnT/TruX} \\
  \textit{University of Luxembourg} \\
  Kirchberg, Luxembourg \\
  alioune.diallo@uni.lu \\
   \And
    Anta DIOP\\
    \textit{UCAD/ESP} \\
    \textit{Université Cheikh Anta Diop} \\
    Dakar, Senegal \\
    antadiop1@esp.sn \\
    \And
    Abdoul Kader KABORE\\
   \textit{SnT/TruX} \\
    \textit{University of Luxemboug} \\
    Kirchberg, Luxembourg \\
    abdoulkader.kabore@uni.lu\\
    \And
    Jordan SAMHI \\
  \textit{SnT/TruX}\\
  \textit{University of Luxemboug} \\
  Kirchberg, Luxembourg \\
    jordan.samhi@uni.lu \\
  \And
    Aleksandr PILGUN\\
    \textit{SnT/TruX} \\
    \textit{University of Luxemboug} \\
    Kirchberg, Luxembourg \\
     aleksandr.pilgun@uni.lu \\
     \And
     Tegawendé F. BISSYANDE\\
        \textit{SnT/TruX} \\
        \textit{University of Luxemboug} \\
        Kirchberg, Luxembourg \\
        tegawende.bissyande@uni.lu
      \And
        Jacque KLEIN\\
        \textit{SnT/TruX} \\
        \textit{University of Luxemboug} \\
        Kirchberg, Luxembourg \\
        jacques.klein@uni.lu
}

\maketitle

\begin{abstract}
Android's open-source nature facilitates widespread smartphone accessibility, particularly in price-sensitive markets.
System and vendor applications that come pre-installed on budget Android devices frequently operate with elevated privileges, yet they receive limited independent examination. 
To address this gap, we developed a framework that extracts APKs from physical devices and applies static analysis to identify privacy and security issues in embedded software.
Our study examined 1,544 APKs collected from seven African smartphones.
The analysis revealed that 145 applications (9\%) disclose sensitive data, 249 (16\%) expose critical components without sufficient safeguards, and many present additional risks: 226 execute privileged or dangerous commands, 79 interact with SMS messages (read, send, or delete), and 33 perform silent installation operations. 
We also uncovered a vendor-supplied package that appears to transmit device identifiers and location details to an external third party. 
These results demonstrate that pre-installed applications on widely distributed low-cost devices represent a significant and underexplored threat to user security and privacy.
\\
\end{abstract}

\begin{center}
\end{center}

\section{Introduction}
 The Android operating system has enabled affordable smartphones for millions of people worldwide. Devices typically ship with manufacturer- or vendor-installed system and third-party apps, which can become a distribution vector for malware and privacy-invasive functionality, especially in developing regions~\cite{kaspersky}. Low-cost smartphones remain widely used across Africa~\cite{trustonic} and have helped reduce the digital divide by expanding access to services~\cite{alfred,dan}. For example, roughly 20–22 million people in South Africa use smartphones, accounting for one third of the population~\cite{trustonic}. Feature phones and inexpensive devices are still popular in Africa, which preserves the market opportunity for low-cost Android devices expansion. Although low-cost Android initiatives have improved device accessibility~\cite{barton}, the pre-installation process can be abused to embed harmful apps that reach large user populations.

 Multiple reports document dangerous apps shipped with devices~\cite{bobe,fakeApp,DangApp,AndMal,10.1145/2590296.2590313,builtMal}. These apps have been observed exfiltrating personal data to remote servers~\cite{bobe}, harvesting financial credentials via clipboard monitoring~\cite{fakeApp}, performing silent installations~\cite{10.1145/2590296.2590313}, or enrolling users in paid services without consent~\cite{builtMal}. Several of the reported apps appear specifically on low-cost devices, including phones sold in Africa~\cite{bobe,builtMal}.

A few studies collected firmware images from vendor websites and online forums, extracting pre-installed apps for analysis~\cite {10.1145/2590296.2590313,251554,9793923,FirmwareDroid}. The extracted apps are then examined for suspicious permissions, misconfigurations in manifest files~\cite{10.1145/2590296.2590313,9793923,FirmwareDroid}, malware~\cite{10.1145/2590296.2590313}, and data leaks~\cite{251554,9152633,9519485}. 
Other studies have directly extracted pre-installed apps from physical devices~\cite{9152633,9519485}, allowing for a more comprehensive assessment across various Android brands. Yet, research on pre-installed applications remains limited since Android devices manufacturers are expected to strictly follow compliance guidelines enforced by Google~\cite{androidcertified}. Unlike globally known brands, Android devices sold in African markets are often very low‑priced. This raises concerns about the trade‑offs manufacturers make between cost and security.

We studied seven low-cost device models (Infinix, Tecno, itel) — all produced by TRANSSION and together accounting for 51\% of the African smartphone market in 2023~\cite{markSHared}. We extracted a dataset of \num{1544} pre-installed APKs for analysis.
Pre-installed apps on these devices have received little systematic analysis despite their prevalence. The most notable work is by Elsabagh et al.~\cite{251554} who examined pre-installed apps from Infinix and Tecno, focusing on privilege escalation. To the best of our knowledge, no prior work has systematically analyzed pre-installed apps across these low cost Android devices commonly sold in Africa.
To explore this gap we developed \texttt{\textit{PiPLAnD}}, a pipeline that extracts APKs from physical devices and applies taint-based leak detection, pattern-driven behavior scanning, and manifest/component inspection to find data leaks, insecurely exported components, and suspicious behaviors.

Our analysis uncovered multiple suspicious pre-installed apps across the tested devices. A substantial fraction of these apps leak sensitive information (for example, IMEI, IMSI, location) and exhibit manifest misconfigurations that expose components without adequate protection. In a notable case, \texttt{\textit{PiPLAnD}} identified the package \textit{com.transsion.statisticalsales}, which was not flagged by VirusTotal. These findings underscore the need for independent audits of pre-installed software on low-cost devices in developing regions.

Below we summarize our main contributions:
\begin{itemize}
    \item[\wcircle{1}] We present \texttt{\textit{PiPLAnD}}, a pipeline to systematically inspect pre-installed apps from Android devices.
    \item[\wcircle{2}] We collected a dataset of \num{1544} pre-installed APKs from seven low-cost device models (Infinix, Tecno, itel).
    \item[\wcircle{3}] We identify manifest misconfigurations: 249 app versions (16\%) export sensitive components without adequate protection.
    \item[\wcircle{4}] We quantify sensitive-data leakage: 145 apps (9\%) leak identifiers or location data.
    \item[\wcircle{5}] We document widespread suspicious behaviors in pre-installed apps.
\end{itemize}

\section{Background}
This short background defines terms used in the paper and clarifies our measurement scope.

\textbf{Android firmware and pre-installed apps.} Firmware boots device hardware and the Android OS~\cite{tLacoma}. Devices ship with \textit{system apps} and \textit{manufacturer-supplied} pre-installed apps~\cite{RolandH}. \textit{System apps} can run with elevated privileges or reside in privileged locations (e.g., \texttt{/system/priv-app}). Some devices in our dataset use Android Go, a lightweight Android configuration for entry-level hardware. Android Go was designed to optimise its running on low-end devices, however the security implications of this optimisation are not fully understood~\cite{androidgo}.

\textbf{Exported components and misconfigurations.} Apps declare components in AndroidManifest.xml. A component is "exported" if other apps can start or bind to it (via \texttt{android:exported} or an \texttt{intent-filter}). Exported components without proper permissions or access checks form an attack surface and are treated here as security misconfigurations~\cite{cweExpComp,exported}.

\textbf{PII, sources, sinks, and leaks.} PII includes identifiers and user data (IMEI, IMSI, phone number, location)~\cite{pii}. A SOURCE is a method that reads sensitive data (e.g., \texttt{getLastKnownLocation()}); a SINK is a method that can exfiltrate data (network send, SMS, world-readable storage). When an app allows to get this data and send it outside the app itself, in this case, we talk about data leakage (or data leak). Static taint analysis tools like FlowDroid are well known to detect leaks in Android apps~\cite{10.1145/2594291.2594299}.

\textbf{Low-cost device.}
In this study, we often use the terms "low-cost", "low-end", "cheap", and "affordable" when talking about devices that are not expensive but affordable, and specifically targeting devices primarily used in Africa.

\section{Low-cost Android devices in Africa}
\label{devices}

Upon careful investigation of the African mobile device market we identified three popular brands shipping Android devices under 100 US dollars price -- Infinix, itel, and Tecno. All these devices run Android Go Edition. We acquired 7 devices in total from these brands for our study.

\subsection{Android Go Edition}

Android Go Edition is a lightweight configuration of standard Android designed for entry-level devices with limited memory ($\leq$ 2\,GB) and storage~\cite{androidgo}. Android Go ships with resource-optimized "Go" versions of Google apps, but users can still install apps from the Play Store. Android Go may receive updates less frequently than standard Android~\cite{goVSstandard,androidgo}. To reduce resource consumption, Android Go disables several features by default~\cite{androidgo}:
\begin{itemize}
    \item Picture-in-picture support;
    \item \texttt{SYSTEM\_ALERT\_WINDOW} permission (display over other apps);
    \item Split-screen / multi-window;
    \item Live wallpapers;
    \item Multi-display;
    \item Launcher shortcuts (deep shortcuts);
    \item Reduced maximum width/height for images in remote views;
    \item VR mode.
\end{itemize}

Android Go is not necessarily less secure than standard Android. In fact, users may gain security benefits from the removal of the \texttt{SYSTEM\_ALERT\_WINDOW} permission, which has often been abused by malicious apps~\cite{fratantonio2017cloak}. We did not find literature that evaluates the security implications of the other optimizations introduced in Android Go. However, Android Go devices typically receive fewer updates and have a shorter support lifecycle, which can leave users without critical security patches~\cite{acar202450}.

Many pre-installed apps on some device brands are not available on Google Play. They may come from unverified third-party vendors that may potentially distribute malicious apps to the end users.

\subsection{Devices and Pre-installed apps}

\begin{figure}[!h]
\centerline{\includegraphics[width=0.45\linewidth]{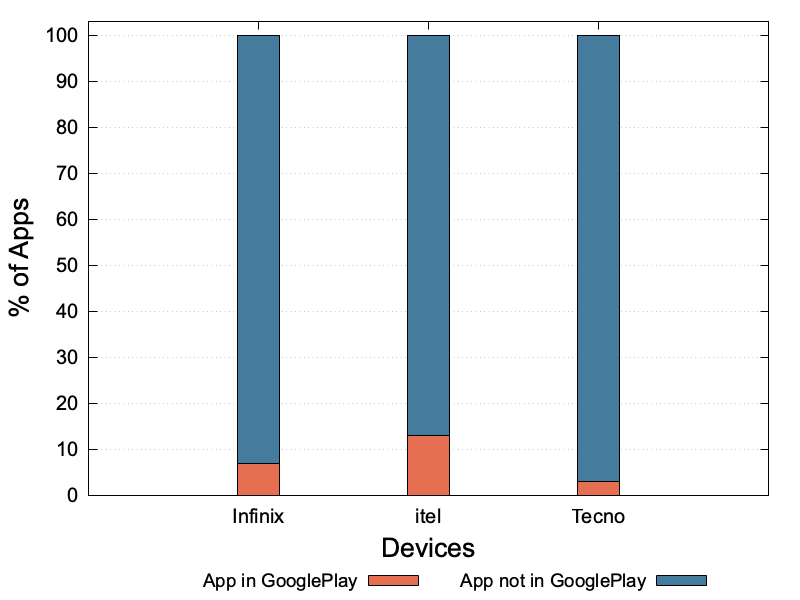}}
\caption{App present in Google Play vs. App not present in Google Play}
\label{fig:AppGP}
\end{figure}

From our seven devices (Infinix, itel, and Tecno) we extracted \num{1544} pre-installed APK files, including both system and third-party apps, which form the dataset used in this study.
Most of the pre-installed apps extracted cannot be found in the Google Play Store, as shown in Figure~\ref{fig:AppGP}.
Several brands provide their own app stores. In particular, the Palm Store is reported to be the official app distribution platform for Infinix, Tecno, and itel~\cite{palmstore}. Palm Store is preinstalled on the devices we examined and allows users to install, uninstall, and update apps. According to the provider, the store performs automated and manual security checks, compatibility testing, and content-compliance monitoring~\cite{palmstore}.

\begin{figure}[!h]
\centerline{\includegraphics[width=0.5\linewidth]{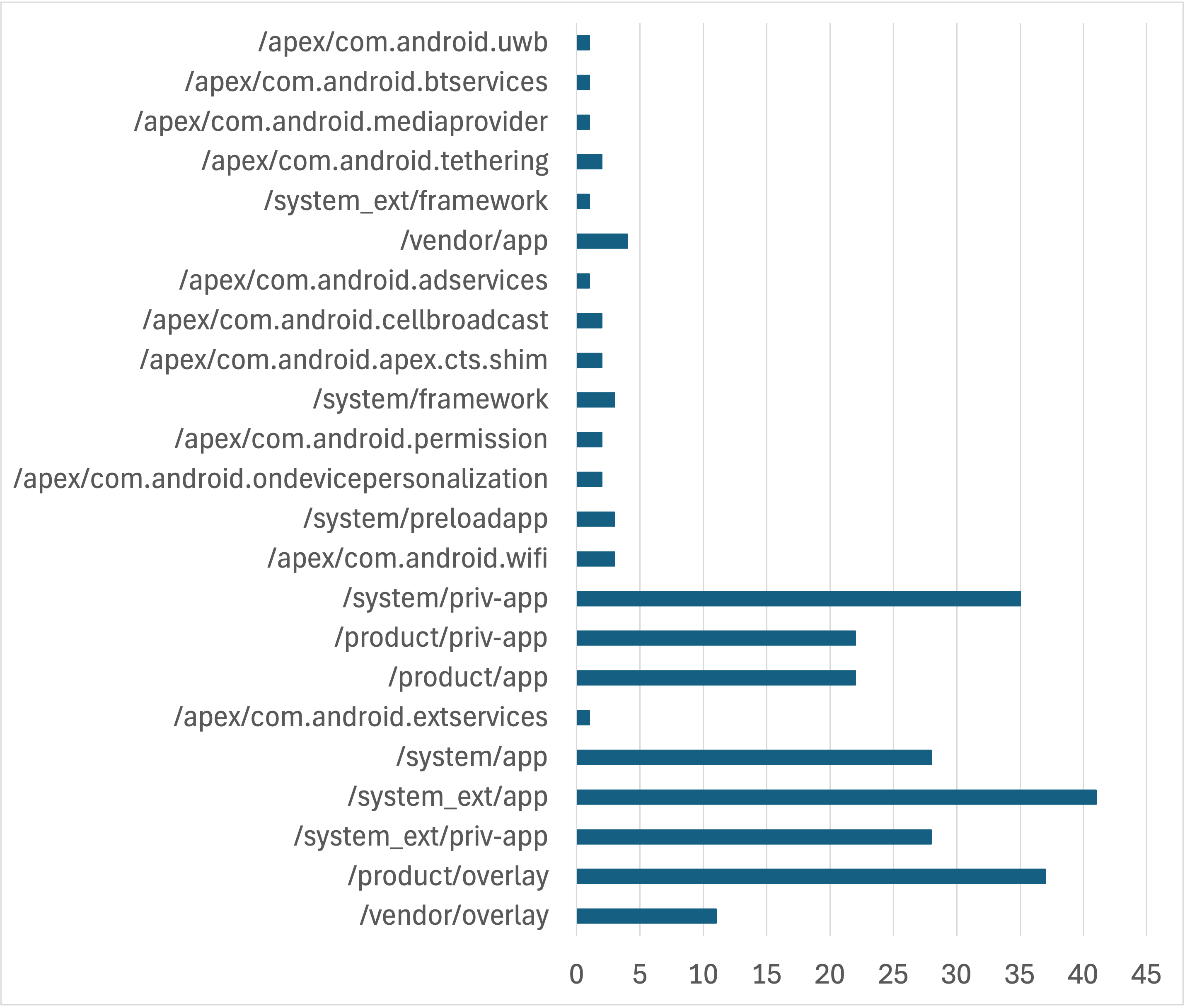}}
\caption{Location folders of the system apps in Infinix}
\label{fig:locFolder}
\end{figure}

System apps are signed by multiple different certificate authorities, and the dominant signer varies by device. For example, Table~\ref{tab:Sig} shows that on the Infinix SMART8, 68\% of system apps are signed by Infinix; 20\% by Google; 4\% by Transsion; 2\% by Tecno; 1\% by Facebook; 1\% with the default AOSP certificate; and 4\% by other authorities.
\begin{table}[!h]
\caption{System apps grouped by certificate authority for some devices.}
\begin{center}
\begin{adjustbox}{width=0.5\linewidth}
\begin{tabular}{|c|c|c|c|}
\hline
\textbf{Signed by} & \textbf{\textit{Infinix SMART8}}& \textbf{\textit{itel A50}}& \textbf{\textit{Tecno POP8}} \\
\hline
Infinix & 68\% & 0\% & 0\% \\
\hline
itel & 0\%& 71\% & 0\% \\
\hline
Tecno & 2\% & 1\% & 68\% \\
\hline
Transsion & 4\% & 2\% & 3\% \\
\hline
Google & 20\% & 22\% & 21\% \\
\hline
Facebook & 1\% & 1\% & 2\% \\
\hline
Default & 1\% & 1\% & 1\% \\
\hline
SW & 0\% & 0\% & 2\% \\
\hline
Others & 4\% & 2\% & 3\%  \\
\hline
\end{tabular}
\end{adjustbox}
\label{tab:Sig}
\end{center}
\end{table}
System apps are usually found on \textit{/system/app} and \textit{/system/priv-app} folders.
However, when extracting the APK files, we have found that the apps have been distributed not only in these two folders, but in several others as well.
Figure~\ref{fig:locFolder} shows an example of the folders on Infinix.
Many apps for some brands have been pre-installed based on the regional needs. For example, the supply chain of Transsion devices suggests that Transsion works with partners in Africa, which could pre-install apps and services tailored for the regional needs~\cite{regneed}.
Since many steps exist to pre-install apps, this could be an entry point for introducing malicious apps on these devices~\cite{preinstMal}.

\section{Methodology}
\label{method}

\subsection{Research questions}
Understanding the potential risks of pre-installed apps is essential.
To guide our research, we have established the following key questions.

\textbf{RQ1: To what extent do pre-installed apps leak sensitive data on low-cost devices?}
This question focuses on checking whether pre-installed apps perform activities, such as leaking sensitive data via the Internet or using other ways.
This helps us to understand not just if these apps pose risks, but also how they do so.


\textbf{RQ2: To what extent do pre-installed apps exhibit suspicious behaviors on low-cost devices?}
The aim of this question is to determine how widespread suspicious pre-installed apps are on affordable devices. Although previous research has shown that some of these apps may contain harmful code~\cite{10.1145/2590296.2590313}, a more comprehensive assessment is needed to quantify the extent of the problem in different manufacturers and regions.Through this question, we will explore the devices to identify apps having suspicious behaviors (including malware) as well as those using malicious URLs.

\textbf{RQ3: How prevalent are security misconfigurations in the manifest files of pre-installed apps on low-cost devices?}
Through this question, we explore apps to identify those exposing sensitive data across the exported components. 


By addressing these research questions, our study provides a clearer picture of the security landscape around pre-installed apps on low-cost mobile devices and highlights potential areas where better security measures are needed.

\subsection{PiPLAnD Design}
\begin{figure}
\centerline{\includegraphics[width=0.7\linewidth]{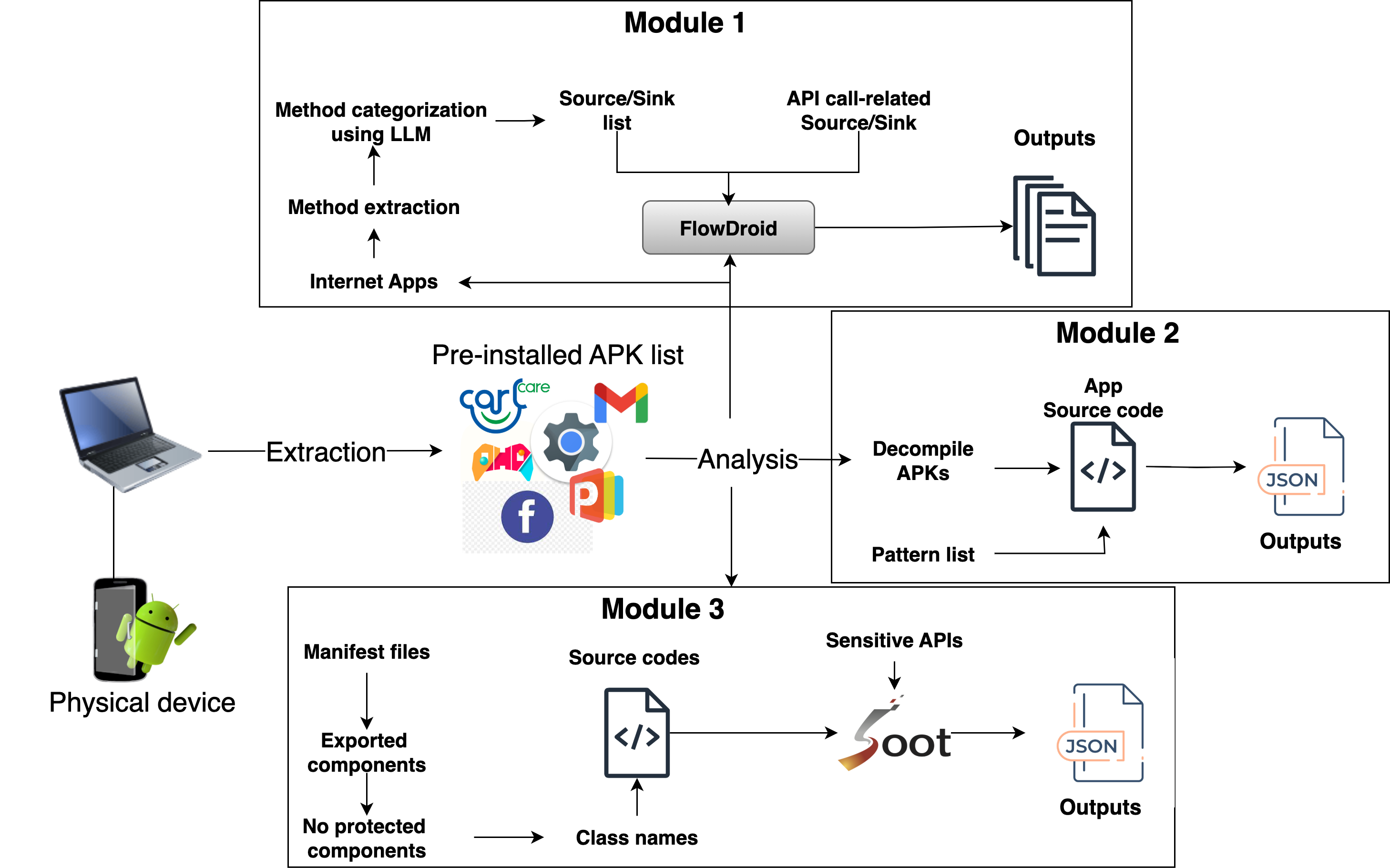}}
\caption{Overview of \texttt{\textit{PiPLAnD}}'s workflow.}
\label{fig:PiPLAnD}
\end{figure}

This section outlines the workflow of \texttt{\textit{PiPLAnD}}, a pipeline designed to inspect pre-installed apps using static analysis approaches. 
Figure~\ref{fig:PiPLAnD} gives an overview of the pipeline.
The source code of \texttt{\textit{PiPLAnD}} is available on our github repository\footnote{PiPLAnD source code: \url{https://anonymous.4open.science/r/PiPLAnD-9C86}}.
We present the details of the workflow of each module in the following sections.

\textbf{Data collection.}
We have used seven (7) low-cost Android devices in this study.
\texttt{\textit{PiPLAnD}} first automatically extracts the pre-installed apps from a physical device using ADB commands when we plug the device into the computer. 
After extraction, we got a dataset of \num{1544} APK files from the overall devices.
This dataset includes system and third-party apps.

\textbf{Module 1: Data leak detection.}\\
We integrate FlowDroid into \texttt{\textit{PiPLAnD}} to identify pre-installed apps that leak sensitive data.
FlowDroid is based on a list of sources and sinks for the detection.
This list contains Android API calls and Java methods.  
Thus, by default, it cannot detect data leaked from methods other than Android API calls and Java methods.
During our study, we have noticed that there are pre-installed apps that use custom methods from third-party libraries to send data over the Internet.
Because of that, we have identified apps that access to Internet by looking for \textit{android.permission.INTERNET} permission on their manifest files.
For these apps, we extracted all their methods that we have given to LLMs for source and sink categorization.
The resulting list is used, in combination with the default list of sources and sinks, to identify Internet apps leaking sensitive data over the Internet.
The default Source and Sink list is also used for detecting other types of data leakage in other apps.

\textbf{Module 2: Behavior analysis.}\\
The behavior analysis focuses on pattern detection.
More specifically, we looked for patterns such as \textit{"pm install"}, \textit{"installPackage"}, etc., to detect installation behaviors, \textit{"logcat"} for apps collecting log data, \textit{"content://sms", "delete", "sendTextMessage", "Telephony.SMS\_RECEIVED"} for accessing SMS, deleting, sending, and listening to received SMS, etc.
Table~\ref{tab:behavPatterns} shows the full list of the patterns we used.
\begin{table}[!h]
    \centering
    \caption{The list of patterns}
    \begin{adjustbox}{width=0.5\linewidth}
    \begin{tabular}{|l|l|}
        \hline
         \textbf{Behaviors} & \textbf{Patterns} \\
         \hline
          & content://sms \\
          & delete \\
          Access / Delete / Send SMS & sendTextMessage \\
          & Telephony.SMS\_RECEIVED \\
          & Telephony.Sms \\
         \hline
         & "am start " \\
          & "chmod " \\
          Dangerous commands & "su "  \\
          & "sudo " \\
          & "rm -rf " \\
         \hline
          Log collection & logcat  \\
         \hline
          & "pm install " \\
         Installation indicators & installPackage \\
          &   \\
         \hline
    \end{tabular}
    \end{adjustbox}
    \label{tab:behavPatterns}
\end{table}
\texttt{\textit{PiPLAnD}} helps with this analysis by decompiling the APK file using the Androguard framework and by checking for patterns that match our list in the app code.
We only consider these patterns since they are the most obvious and are based on the existing literature, as well.
Then, it reports the findings in a JSON file.
This file is used for a manual analysis to confirm the behavior of the app based on the patterns detected in its code.

\textbf{Module 3: Security misconfigurations on Manifest files.}\\
This module allows for the identification of security misconfigurations in the Android manifest files.
In this study, we only focus on identifying components exported that allow access to sensitive information.
Android allows restricting access to an exported component by using permissions.
We first extracted exported components from the manifest files for each app.
Then, we filter out and keep those who did not have permission to protect and restrict their access.
This gives us a list of class names with their full path in the app code.
Besides that, we have constructed a list of sensitive API calls based on the source and sink list from FlowDroid, and from the Androwarn study~\cite{debize2012androwarn}, in which authors look for sensitive data accessed.
We explored the list containing the classes of the exported components, and for each class, we analysed it using the SOOT framework~\cite{lam2011soot}.
For the analysis with SOOT, we used the sensitive API list and looked for each of them in the code of the corresponding class to know whether this component allows access to sensitive data.
If we did not find the presence of a sensitive API in the code, our analysis program constructs a call graph (CG) of the corresponding class to search whether the sensitive API is used in a method called in the code of the exported component.
The results are put into a JSON file, in which we have the corresponding class, the sensitive API, and the method where we found this API.

\section{Analysis Results}
\label{result}

In this section, we present the findings from seven (7) identified affordable devices that are primarily used in Africa. 
The main goal of this study is to evaluate the security of the pre-installed apps and to investigate the prevalence of suspicious one on low-cost devices.
Our analysis has revealed interesting findings.
Table~\ref{tabResults} summarizes these findings. 
This section shows the results of the analysis by answering the different research questions.
\begin{table*}[!htbp]
\caption{Summary of the results.}
\begin{center}
\begin{adjustbox}{width=0.95\linewidth}
\begin{tabular}{ll|c|}
\cline{1-3}
\multicolumn{1}{ |l| } {\textbf{Analysis Modules}} & \textbf{Behaviors}& \textbf{\# of apps (\%)} \\ \cline{1-3}
\multicolumn{1}{ |l| }{Exported sensitive components} & Components that allow to access sensitive data& \footnotesize{249 (16\%)} \\ \cline{1-3}
\multicolumn{1}{ |l| } {Leak of sensitive data}  & Leak of sensitive data & \footnotesize{145 (9\%)} \\ \cline{1-3}
\multicolumn{1}{ |l  }{\multirow{2}{*}{} } &
\multicolumn{1}{ |l| }{Dangerous commands} & \footnotesize{226 (15\%)} \\ \cline{2-3}
\multicolumn{1}{ |l  }{}                        &
\multicolumn{1}{ |l| }{Log collection} & \footnotesize{10 (0.7\%)} \\ \cline{2-3}
\multicolumn{1}{ |l  }{Suspicious behaviors}                        &
\multicolumn{1}{ |l| }{Silent installation behaviors} & \footnotesize{33 (2\%)} \\ \cline{2-3}
\multicolumn{1}{ |l  }{}                        &
\multicolumn{1}{ |l| }{Access / Send / Delete SMS} & \footnotesize{79 (5\%)} \\ \cline{1-3}
\end{tabular}
\end{adjustbox}
\label{tabResults}
\end{center}
\end{table*}

\subsection{Apps leaking sensitive data}

Mobile apps often handle sensitive data on the device. 
With the privileges that pre-installed apps have, they can perform harmful activities, including leaking this data.
Our analysis, consisting of identifying apps that leak sensitive data, has identified several pre-installed apps on seven 7 different devices leaking data, such as Mobile Country Code (MCC), user location (longitude and latitude), device info, International Mobile Subscriber Identity (IMSI), IMEI (International Mobile Equipment Identity), etc.
Sensitive data is leaked in different ways, such as in SharedPreferences, in logs, in Intents, as well as on the network.
Overall, we have found around 9\% of the pre-installed apps on the devices leaking sensitive data, which represent 145 pre-installed apps. 
As an example, the app (\textit{com.transsion.statisticalsales}) sends sensitive information to a remote host by using third-party libraries with customized methods. 
As illustrated in Listing~\ref{listing:5}, the app collects location info, IMSI, IMEI, phone version, etc., and sends them to a remote server.
Pre-installed apps also leak data using other ways, such as SharedPreferences storage, device logs, etc.
These practices expose the user to many security and privacy problems. 
\begin{listing*}[!h]
\caption{App sending sensitive data to a remote server}
\label{listing:5}
\begin{minted}[fontsize=\scriptsize, baselinestretch=0.9, breaklines, breakanywhere]{java}
[...]
public class SSHttpClient {
    public static final String BASE_URLFLAG = "PCHttpClient";
    private static final String DEFAULT_BASE_SERVER = "https://asv.transsion.com:443/SaleStatistics/sendsale/sendSale";
    private static final String DEFAULT_INDIA_SERVER = "https://asvin.transsion.com:8080/SaleStatistics/sendsale/sendSale";
    protected static final String TAG = "SSHttpClient_";
    [...]
    public void RegisterInformation(final HttpCallback<HttprequestResult> httpCallback, boolean z) {
        RequestParams requestParams = new RequestParams();
        requestParams.put("ua", this.requestInfo.getUa());
        requestParams.put("screen", this.requestInfo.getScreen());
        requestParams.put("imsi", this.requestInfo.getImsi());
        requestParams.put("imei", this.requestInfo.getImei());
        requestParams.put("phone_version", this.requestInfo.getPhone_version());
        requestParams.put("platform", this.requestInfo.getPlatform());
        requestParams.put("device", this.requestInfo.getDevice());
        requestParams.put("lang", this.requestInfo.getLang());
        requestParams.put("timeStamp", this.requestInfo.getTimeStamp());
        requestParams.put("auth", this.requestInfo.getAuth());
        requestParams.put("lat", this.requestInfo.getLat());
        requestParams.put("lng", this.requestInfo.getLng());
        requestParams.put("client_type", this.requestInfo.getClient_type());
        requestParams.put("phone", this.requestInfo.getPhone());
        requestParams.put("client_version", this.requestInfo.getClient_version());
        requestParams.put("lac", this.requestInfo.getCELL_LAC());
        requestParams.put("cid", this.requestInfo.getCELL_CID());
        mClient.post(z ? DEFAULT_INDIA_SERVER : DEFAULT_BASE_SERVER, requestParams, new AsyncHttpResponseHandler() { 
            [...]
        });
    }
}
\end{minted}
\end{listing*}
\begin{figure}[H]
\begin{tcolorbox}[
    colframe=brown!30!black, colback=red!2, 
    title=Answer to RQ1,
    fonttitle=\bfseries,
]
The results show that several pre-installed apps on the three devices exhibit harmful activities such as leaking sensitive data, exposing the users to severe risks.
\end{tcolorbox}
\end{figure}



\subsection{Suspicious behaviors on pre-installed apps}
Since mobile devices come with pre-installed apps, these may contain harmful code that can compromise the security of users~\cite{10.1145/2590296.2590313}.
We have looked for apps having suspicious behaviors.
Several pre-installed malware have been identified to have silent installation behaviors~\cite{10.1145/2590296.2590313}. Others steal sensitive data using different ways~\cite{10.1145/2590296.2590313,contDev,trojanAnd}.
To identify these kinds of malware, we focus the analysis in detecting patterns.
When a pre-installed app declares the \textit{INSTALL\_PACKAGES} permission, it has the ability to install an app without the user's knowledge.
It can use \textit{android.content.pm.PackageManager.installPackage()} or \textit{Runtime.exec( 'pm install')} to silently install the app~\cite{10.1145/2590296.2590313}.
Our analysis has revealed 33 pre-installed apps having this silent installation behavior on the overall devices.
Several pre-installed apps have access to the SMS provider \textit{content://sms}.
Some of them delete SMS or declare the \textit{SEND\_SMS} and \textit{RECEIVE\_SMS} permissions with the broadcast action \textit{Telephony.SMS\_RECEIVED}, allowing them to listen and send messages. We have found around 79 apps pre-installed on these low-cost devices, sending, deleting, and/or reading SMS contents.
Furthermore, at least 10 pre-installed apps can access and collect the logcat content. Since they have déclared the  \textit{READ\_LOGS} permission, they can access the overall logcat content, including logs from other applications.
In addition to this, we have found several apps that execute dangerous commands (226 apps). 

\begin{figure}[h!]
\begin{tcolorbox}[
    colframe=brown!30!black, colback=red!2, 
    title=Answer to RQ2,
    fonttitle=\bfseries,
]
Low-cost Android devices ship pre-installed apps with suspicious behaviors, including sending/deleting/reading SMS, executing dangerous commands, accessing overall logcat memory, and having silent installation behaviors.
\end{tcolorbox}
\end{figure}

\subsection{Security misconfigurations on the manifest files}
\textbf{Exported sensitive components.}
Android apps often export components, such as activities, services, receivers, and providers, explicitly by setting \textit{android:exported="true"} or implicitly by declaring an \textit{intent-filter} in the manifest file.
When it is the case, the app allows other apps to launch the component~\cite{exported}.
The access to exported components is often restricted using permissions~\cite{component}. If there is permission for an exported component, the app that wants to launch it should declare this permission.
If an app exports a component without properly enforcing permission, any app could launch it or access sensitive data it contains~\cite{cweExpComp}.
When a component allows access to sensitive data, we call it a sensitive component.
Our analysis has revealed that several pre-installed apps have sensitive components exported.
As illustrated in Table~\ref{tabResults}, we have found around 16\% of the pre-installed apps, representing 249 different app versions, having exported sensitive components on the low-cost devices, without any restriction or protection mechanism.
It means that these devices embed apps that allow other apps to potentially access sensitive information, exposing the user to security and privacy problems.
For example, we have found a pre-installed app (\textit{com.transsion.carlcare}) having this method (Listing~\ref{listing:1}), from an exported activity (\textit{com.transsion.carlcare.WarrantyCardActivity}).
This method allows access to location information (lines 13 and 14).
In the Android manifest file of the app, this component is clearly exported; however, it is protected by no mechanism, facilitating its easy access and its potential exploitation.
Another example shows an exported ContentProvider (\textit{com.sprd.providers.photos.SpecialTypesProvider}) found in the app (\textit{com.sprd.providers.photos}) that allows access to media files, contained in an external storage, from a URI (Listing~\ref{listing:3}, line 10). 
This exported component does not have a mechanism that restricts its access by other apps.

\begin{listing}[!h]
\caption{Exported activity accessing sensitive information}
\label{listing:1}
\begin{minted}[fontsize=\scriptsize]{java}
public void R1() {
    [...]
    HashMap<String, String> map = new HashMap<>();
    if (TextUtils.isEmpty(this.f15263g0)) {
        map.put("imei", listA.get(0));
    } else {
        map.put("imei", this.f15263g0);
    }
    map.put("imsi", wd.c.g());
    map.put("lang", getResources().getConfiguration().locale.
                                        toString());
    if (this.f15282z0 != null) {
        map.put("lat", this.f15282z0.getLatitude() + "");
        map.put("lng", this.f15282z0.getLongitude() + "");
    }
    map.put("phone_version", a2());
    map.put("screen", getResources().getDisplayMetrics().widthPixels + "*" + getResources().getDisplayMetrics().heightPixels);
    map.put("ua", Build.BRAND + "-" + Build.MODEL);
    [...]
}
\end{minted}
\end{listing}
\begin{listing}[!h]
\caption{Exported ContentProvider accessing external storage files}
\label{listing:3}
\begin{minted}[fontsize=\tiny]{java}
[...]
public final class SpecialTypesProvider extends ContentProvider {
    [...]
    private static final Uri EXTERNAL_CONTENT_URI = MediaStore.Files.getContentUri("external");
    private static final String[] SPECIAL_TYPE_PROJECTION = {"_data", "owner_package_name"};
    [...]
    private int getCameraType(long j) {
        Log.d(TAG, "mediaStoreId = " + j);
        boolean z = true;
        Cursor cursorQuery = getContext().getContentResolver().query( EXTERNAL_CONTENT_URI, SPECIAL_TYPE_PROJECTION, "_id=?", new String[]{String.valueOf(j)}, null);
        if (cursorQuery != null) {
            try {
                if (cursorQuery.moveToFirst()) {
                    String string = cursorQuery.getString(0);
                    Log.d(TAG, "mediaPath = " + string);
                    String string2 = cursorQuery.getString(1);
                    if (!GOOGLE_PHOTOS_PACKAGE_NAME.equals( string2) && !GOOGLE_GALLERY_PACKAGE_NAME.equals( string2)) {
                        z = false;
                    }
                    Log.d(TAG, "ownerPackageName = " + string2 + ", isGoogleCreatedImage = " + z);
                    try {
                        ExifInterface exifInterface = new ExifInterface();
                        exifInterface.readExif(string);
                        iIntValue = z ? 0 : exifInterface.getTagIntValue( ExifInterface.TAG_CAMERATYPE_IFD). intValue();
                        Log.d(TAG, "getCameraType cameraType = " + iIntValue);
                    } catch (Exception e) {
                        Log.d(TAG, "Exception occurs, mediaPath = " + string + ". ex = " + e);
                    }
                }
            } finally {
                if (cursorQuery != null) {
                    cursorQuery.close();
                }
            }
        }
        return iIntValue;
    }
    [...]
}
\end{minted}
\end{listing}
\begin{figure}[!h]
\begin{tcolorbox}[
    colframe=brown!30!black, colback=red!2, 
    title=Answer to RQ3,
    fonttitle=\bfseries,
]
The results have revealed several pre-installed apps exporting sensitive components, including activities, services, content providers, and receivers, on low-cost devices. This potentially puts users at risk, as their data could be accessed by third parties.
\end{tcolorbox}
\end{figure}

\section{Discussions}
\label{discuss}
In this section, we discuss the results from the analysis of the devices. 

\textbf{Sensitive data leakage.}
During our analysis, we have found several pre-installed apps leaking sensitive data.
Several of them send the data to a remote host by using Android API calls.
Others use third-party libraries that use customized methods to avoid existing detection (e.g, \textit{com.transsion.statisticalsales}).
We have considered these customized methods and added them to FlowDroid, and we have detected apps sending sensitive data over the Internet with these methods.
Considered as malware by some blog posts~\cite {bobe}, the app (\textit{com.transsion. statisticalsales}) is not detectable by malware scanners such as VirusTotal.
It silently collects data (phone version, IMSI, user location, CID (Cell Tower ID), LAC (Location Area Code), etc.) and sends it to a remote host. This suspicious behavior compromises users' security and privacy.
Several other apps get and store sensitive data in internal storage, such as SharedPreferences, or log it in the logcat memory. 
These leaked data can be used by malicious actors for user profiling, user tracking~\cite{leakApp}, or accessed by other apps since pre-installed apps can share each other data when they have the same \textit{sharedUserId} and have signed with the same  certificate~\cite{10.1145/2590296.2590313}.

\textbf{Suspicious behaviors.}
We have found several pre-installed apps having suspicious behaviors, including silent installation behavior, send/delete SMS, etc.
These are done without the user's knowledge and may have negative consequences. 
Indeed, when an app is able to install another app silently, it may install a malicious app from a malicious remote server or via dynamic code loading.
This technique is often used by malicious actors to infect Android devices and avoid detection~\cite{10.1145/2590296.2590313}.
This is possible only with system apps, since non-system apps cannot declare the required permission.
When a pre-installed app has the ability to send and delete messages without the user's interaction, this can lead to the theft of sensitive data.
Several malware are known to use this technique. 
For instance, a variant of the Triada malware has been found pre-installed in Android devices~\cite{contDev}. 
This malware performs several actions, among which we have enabling premium SMS services, intercepting, sending, and deleting messages.

\textbf{Security misconfiguration.}
Our analysis has revealed several apps exporting components that allow other apps to potentially access sensitive data, without any protection.
The Android framework proposes permission levels, including normal, dangerous, and signature, to protect and restrict access to components~\cite{permsComp}.
The absence of protecting exported components is a known vulnerability from the community~\cite{permsComp,cweExpComp}.
When an app exports a component without any restriction, it allows other apps to access it. From this, a malicious app may access to sensitive resources,  as mentioned in the Common Weakness Enumeration (CWE - 926)~\cite{cweExpComp}.
The app can be victim to an attack named confused deputy attack~\cite{10.5555/2028067.2028089,9152633}, as well as a malicious app can abuse the privileges that these components have to gain unauthorized access to these resources~\cite{permsComp,9043935}.

\section{Related Work}
\label{related}

Zheng et al.~\cite{10.1145/2590296.2590313} presented a tool named DroidRay that extracts statically and dynamically pre-installed apps from 250 firmware downloaded from forums and website. 
The static extraction consists of extracting pre-installed apps directly from the firmware images. As for the dynamic extraction, it consists of flashing the image into a device and then extracting the pre-installed apps using ADB commands.
DroidRay performs pre-installed app analysis and system analysis.
For the pre-installed app analysis,  they extracted the “SharedUserId” attribute from the AndroidManifest.xml and the signature information from the RSA file. Then, they compare them with the default signatures they found from the AOSP. They also analyzed the apps in VirusTotal do detect potential malware, and applied a filter to retain apps having dangerous permissions or silent installation behavior.
For the system analysis, DroidRay performs static and dynamic analysis of the Android firmware by doing a system signature vulnerability detection, a network security analysis, and a privilege escalation vulnerability detection. 
In our proposed pipeline, we leveraged the technique of dynamic extraction to directly extract pre-installed apps from a physical device using ADB commands rather than collecting firmware and flashing them into a device.
We also used the same technique concerning malware detection, consisting of analysing APK files in VirusTotal.

Mitchell et al.~\cite{10.1145/2557547.2557557} designed DexDiff, a system for assessing the security impacts of vendor customization to the official Android system.
DexDiff helps the security analyst, who first retrieves the pre-installed apps and libraries from the phone and then builds their corresponding base binaries from the release branch in AOSP on which the phone is based.  
DexDiff compares each pair of these binaries obtained and evaluates the security impacts of individual modifications.
The bit and only similarity that this study has compared to our approach is that it extracts pre-installed apps from the phone. However, the proposed tool did not automate this process. The apps are supposed to be extracted before using DexDiff.
Contrary to our approach, PiPLAnD automated the process of extraction and analysis.

Elsabagh et al.~\cite{251554} proposed a static analysis tool named FIRMSCOPE, to identify unwanted functionality in pre-installed apps by analyzing the Android firmware.
FIRMSCOPE extracts pre-installed apps from the Android firmware and then performs a taint analysis with context-sensitive, flow-sensitive, field-sensitive, and partially object-sensitive.
Specifically, it focuses exclusively on identifying the increase in privileges.

Gamba et al.~\cite{9152633}  presented a large-scale study of Android pre-installed using crowd-sourcing methods.
In this study, the authors built an Android app, Firmware Scanner, that looks for and extracts pre-installed apps when installed on a device.
This study performs permission analysis using Androguard~\cite{desnos2011androguard}, static analysis leveraging existing tools such as Androwarn, FlowDroid~\cite{10.1145/2594291.2594299}, and Amandroid~\cite{wei2018amandroid}, as well as apktool~\cite{apktool} and Androguard frameworks to identify unwanted behaviors, and traffic analysis using the crowd-sourced Lumen mobile traffic dataset to see app real-world behaviors. 
Compared to this work, we did the same by extracting pre-installed apps from the physical device but using ADB commands rather than an installed app, and we also performed a taint analysis using FlowDroid.

Bl\'azquez et al.~\cite{9519485} proposed FOTA (Firmware-Over-The-Air) Finder to automatically classify a given APK as FOTA or not based on Androguard, using the dataset of pre-installed apps from Firmware Scanner~\cite{9152633}. 
Then, they performed behavior analysis relying on FlowDroid and Amandroid for a taint analysis and a modification of Androwarn~\cite{androwarn} to analyze the use of API calls.
This part of the work is a bit similar to our data leak detection using FlowDroid. However, we considered all the pre-installed apps extracted from devices, and performed also malware detection.

Hou et al.~\cite{9793923} performed a study in which they collected firmware images from vendors, official websites and open source repositories, and CVE data to link them with pre-installed apps. 
In this study, they proposed a tool named AndScanner that automates the extraction of pre-installed apps from firmware images before analyzing them.
AndScanner proposes an analysis of the security patches of the firmware to know if it has been patched in time and if the security issues have been fixed. 
It also performs app analysis by analyzing the pre-installed apps using Androguard to identify misconfiguration in the manifest file and CryptoGuard~\cite{CryptoGuardOSS} to detect cryptography misuses.
Our study is completely different since we did not focus on analyzing the manifest files and cryptography misuses. However, we leveraged on Androguard framework in our pipeline.

More recently, Sutter and Tellenbach~\cite{FirmwareDroid} proposed FirmwareDroid, an automated static analysis tool for pre-installed apps.
This tool automates the process of extracting pre-installed apps from firmware images and their analysis using existing tools including Androgurad and Exodus.
The study identifies the advertising tracker libraries used with Exodus and the permissions pre-installed apps inherited with Androguard.
The authors have integrated 8 open source static analysis tools in FirmwareDroid which could be used for more analysis.

Almost all of these studies are a bit similar since they follow almost the same approach, such as collecting firmware from the Internet, extracting the pre-installed apps, and analyzing them.
Our approach allows the extraction of pre-installed apps from physical devices, the detection of malware and dangerous permissions, the detection of data leakage, and the use of malicious URLs in pre-installed apps. 
Furthermore, our approach is particularly tested on low-cost devices sold in Africa.
However, with this approach, every brand and every Android device can be inspected and analyzed. Consequently, we do not have a limitation related to missing some brands or firmware.

\section{Conclusion}
\label{concl}
This study investigates and analyzes Android pre-installed apps, in particular, those shipped with phones sold in Africa.
For this purpose, we have proposed a pipeline that allows the extraction of APK files from a physical device and inspects them to look for different suspicious behaviors, including pre-installed malware, apps exposing sensitive data, and apps sending personal data to remote hosts.
The pipeline is tested by inspecting three different low-cost Android devices bought in Africa.
The results show interesting findings, such as (1) the leak of sensitive data, (2) sensitive data exposure through exported components and (3) apps having suspicious behaviors.
As future research, we planned to go deeper into these pre-installed apps by analysing the URLs they use to identify malicious ones, the native libraries used, and by looking for further suspicious behaviors.
Our future study will extend the number of low-cost devices and compare the results from the analysis of pre-installed apps with those from European devices, as well.
In addition, we planned to perform a dynamic analysis by implementing a solution that monitors the system log and detects suspicious behaviors using AI models.



\bibliographystyle{splncs04}
\bibliography{sample}

\end{document}